\documentstyle[prb,aps,multicol,epsf]{revtex}
\tighten
\begin{document}
\draft

\title{Suppression of non-Poissonian shot noise 
by Coulomb correlations\\ in ballistic conductors}
\author{O. M. Bulashenko and J. M. Rub\'{\i}}
\address{Departament de F\'{\i}sica Fonamental,
Universitat de Barcelona, Diagonal 647, E-08028 Barcelona, Spain}
\author{V. A. Kochelap}
\address{Department of Theoretical Physics,
Institute of Semiconductor Physics, Kiev 252028, Ukraine}

\date{\today}
\maketitle

\begin{abstract}
We investigate the current injection into a ballistic conductor under 
the space-charge limited regime, when the distribution function of
injected carriers is an arbitrary function of energy $F_c(\varepsilon)$.
The analysis of the coupled kinetic and Poisson equations shows that
the injected current fluctuations may be essentially suppressed
by Coulomb correlations, and the suppression level is determined by
the shape of $F_c(\varepsilon)$.
This is in contrast to the time-averaged quantities: 
the mean current and the spatial profiles are shown to be insensitive 
to $F_c(\varepsilon)$ in the leading-order terms at high biases.
The asymptotic high-bias behavior for the energy resolved shot-noise 
suppression has been found for an arbitrary (non-Poissonian) injection,
which may suggest a new field of investigation on the optimization of
the injection energy profile to achieve the desired noise-suppression level. 
\end{abstract}


\begin{multicols}{2}

\section{Introduction}

Randomness in the transmission of discrete charge carriers in mesoscopic 
conductors leads to the fluctuations of the electric current called shot 
noise. \cite{dejong97,blanter99}
Recently, shot-noise measurements are emerging as an important tool to probe 
carrier interactions in mesoscopic systems. \cite{landauer98}
As interactions between electrons can regulate their motion, this
effect may be detected in the shot-noise reduction, but cannot be deduced
from time-averaged dc measurements.
Usually, the shot-noise level is said to be reduced when its spectral density
is lower in respect to the Poissonian value $S_I^{\rm Poisson}=2qI$,
which is characteristic for transmission of uncorrelated carriers.
(Here, $q$ is the electron charge and $I$ is the mean current.)
The sub-Poisson shot noise could arise due to the Pauli exclusion principle
or Coulomb interactions. The diversity of examples is available from recent 
reviews. \cite{dejong97,blanter99} 

A matter of particular interest is the significance of Coulomb interactions 
in scattering-free or {\em ballistic} conductors.
\cite{landauer96,gonzalez97,prb98,prb00}
This subject is important not only from a fundamental,  
but also from an applied point of view.
Indeed, as the dimensions of practical electronic devices are scaled down,
the ballistic component in carrier motion becomes dominant. \cite{beenakker91}
Then, unavoidable electric charge of carriers and their redistribution across 
the device both give rise to the {\em charge-limited ballistic transport}. 
Based on ballistic transport, a variety of new electronic devices 
is currently discussed in view of future applications to ultralarge scale 
integrated circuits, logic, and memory technology,
\cite{natori94,wernersson97,pikus97,facer99}
and new experimental techniques, like ballistic electron emission 
spectroscopy, have already been realized. \cite{kaiser88,heer99} 
In charge-limited ballistic conductors the shot-noise measurements may become
one of the major tools not only to identify the ballistic transport, but
also to probe carrier interactions and other electronic properties.

In the absence of scattering, the transport and noise properties of ballistic
conductors are determined, to a great extent, by the contacts (emitters).
When the injecting contact is in a local equilibrium and the electron density 
injected into a ballistic conductor is low, the electron gas is nondegenerate 
and described by the Maxwell-Boltzmann distribution function.
In this case, the injected electrons are statistically independent obeying 
the Poissonian statistics. The self-consistent theory of shot-noise 
suppression due to Coulomb interactions for this type of ballistic injection 
has been recently developed. \cite{prb00}
However, in nanoscale devices the injected carriers may be degenerate, 
or income from an emitter with an extremely nonequilibrium distribution,
like in a hot-electron transistor, resonant-tunneling-diode emitter,
superlattice emitter, etc. 
(see, e.g., Refs.\ \onlinecite{levi85,heiblum85,palevski89,xie99,mitin99}).
The incoming carriers may be correlated {\em a priori} and follow 
non-Poissonian statistics.

The main purpose of the present paper is to develop a self-consistent theory 
of shot noise in two-terminal ballistic space-charge-limited conductors 
with an {\em arbitrary} injection energy distribution $F_c(\varepsilon)$,
which would also be valid for {\em any} given correlation properties of 
injected carriers.
To distinguish the pure effect of Coulomb interactions on the shot-noise 
suppression, it will be convenient to measure the noise-suppression level 
in respect to the shot noise of {\em non-Poissonian} flow with disregarded 
Coulomb interactions, rather than to the Poissonian $2qI$ value.
We have derived the analytical formulas that determine the steady-state
and noise characteristics in ballistic conductors under the action of Coulomb 
interactions in the asymptotic limit of high biases.
The time-averaged quantities are found to be insensitive to 
$F_c(\varepsilon)$ in the leading-order terms, giving, in particular, 
the universal Child law for the mean current.
In contrast, the current noise is shown to be crucially dependent on 
$F_c(\varepsilon)$, with the noise suppression (caused by Coulomb 
interactions) different for different injections. 
The derived energy-resolved shot-noise suppression formulas indicate 
the possibility to probe the injection energy profile of a ballistic emitter 
in shot-noise measurements, thereby obtaining an important information
not otherwise available from time-averaged conductance measurements.
On the other hand, that information may help to optimize the injection 
energy distribution to achieve the desired noise-suppression level.

The paper is organized as follows.
In Sec.\ II we introduce the basic equations that describe
the space-charge-limited ballistic transport: the collisionless kinetic
equation coupled self-consistently with the Poisson equation.
The obtained solutions allow us to find the analytical formulas 
for the mean current and the current fluctuation transmission 
expressed through the injection distribution function $F_c(\varepsilon)$.
The shot-noise suppression factor is calculated for some particular
cases in Sec.\ III. 
Finally, Sec.\ IV summarizes the main contributions of the paper,
whereas in the appendix we present mathematical details concerning
the derivation of the self-consistent potential fluctuations.

\section{Transport and noise in space-charge-limited ballistic conductors}

\subsection{The physical model}

Consider a two-terminal semiconductor ballistic sample with plane-parallel
heavily doped contacts at $x$=0 and $x$=$l$.
The structure may be considered as a $n$-$i$-$n$ heterodiode \cite{prb00}
operating under a space-charge-limited current regime in which the current 
is determined by a charge injection from the contacts rather than by 
intrinsic carriers of the ballistic region. 
The applied bias $U$ between the contacts is assumed to be fixed by
a low-impedance external circuit. 
In order to simplify the problem, we assume that due to the large difference 
in the carrier density between the contacts and the sample, and hence in 
the corresponding Debye screening lengths, all the band bending occurs 
in the ballistic base,
and therefore the relative position of the conduction band and the Fermi
level $\varepsilon_c-\varepsilon_F$ does not change in the contacts.
For such a modeling, all of the potential drop takes place exclusively inside
the ballistic base and the contacts are excluded from the consideration.
\cite{gonzalez97,prb98,prb00}
In contrast to Refs.\ \onlinecite{gonzalez97,prb98,prb00}, 
the injected carriers are not restricted to follow a thermal
equilibrium distribution, their distribution is an arbitrary function 
determined by the particular properties of the emitter.
Assuming the transversal size of the conductor sufficiently thick
and high enough electron density, the electrostatic problem may be 
considered in a one-dimensional plane geometry. \cite{prb00} 

\subsection{Distribution function and its fluctuation
in a self-consistent field}

A semiclassical ballistic transport is described by the collisionless 
kinetic equation for the time-dependent distribution function 
$\tilde{F}(x,k_x,t)$ coupled self-consistently with the Poisson equation 
for the electrostatic potential $\tilde{\varphi}(x,t)$,

\begin{eqnarray} \label{vlasns}
\left(\frac{\partial}{\partial t} + 
\frac{\hbar k_x}{m} \frac{\partial}{\partial x} \right.
&+& \left. q \frac{d\tilde{\varphi}}{d x} \frac{\partial}{\hbar\partial k_x}
\right) \tilde{F}(x,k_x,t) = 0, \\ 
\frac{\partial^2\tilde{\varphi}}{\partial x^2} &=& \frac{q}{\kappa} \,
\int  \tilde{F}(x,k_x,t) \frac{\hbar dk_x}{\sqrt{2m}},
\label{pois}
\end{eqnarray}
where $\kappa$ is the dielectric permittivity,
and $m$ the electron effective mass. 
Since during the ballistic motion only the longitudinal momentum $k_x$ 
may vary, we use the electron distribution function averaged over 
the transversal momentum ${\bf k}_{\perp}$ according to 

\begin{equation} \label{fperp}
\tilde{F}(x,k_x,t)= \frac{\sqrt{2m}}{\hbar}
\int \frac{d {\bf k}_{\perp}}{(2\pi)^d} f(x,k_x,{\bf k}_{\perp},t),
\end{equation}
where $d$ is the dimension of a momentum space and 
$f(x,k_x,{\bf k}_{\perp},t)$ 
is the occupation number of a quantum state at the cross section $x$.
The additional multiplication factor $\sqrt{2m}/\hbar$ in the integral 
(\ref{fperp}) is introduced for further normalization convenience.
Under the space-charge-limited transport conditions, the distribution function
$F$ and the space charge in the Poisson equation (\ref{pois}) are determined by
the electrons injected from the contact. 
Due to the stochastic nature of the injection, the distribution function 
$\tilde{F}(x,k_x,t)=F(x,k_x)+\delta F(x,k_x,t)$ and the potential
$\tilde{\varphi}(x,t)=\varphi(x)+\delta\varphi(x,t)$ fluctuate in time
around their time-averaged values.
The nonuniform distribution of the injected carriers
creates the potential minimum $\tilde{\varphi}_m(t)$ at a position 
$x=x_m$, which also fluctuates.
It is the potential minimum fluctuations, that leads to the suppression
of the injected current fluctuations. \cite{prb00}
We assume that the applied bias is much larger than the characteristic
energy spreading of injected electrons, so that the current injection 
from the second (receiving) contact is negligible.
Another assumption is $U_m \ll U < U_{cr}$, where $U_m\equiv -\varphi_m$,
and $U_{cr}$ is the bias at which the potential barrier vanishes.
\cite{remark2}
This assumption may be fulfilled under the condition of a strong screening
that corresponds to the so-called ``virtual cathode'' approximation,
when the potential minimum is so close to the contact, that one can disregard
the region between the contact and the minimum. \cite{prb00}
In this limit, only those electrons that are able to pass over 
the fluctuating barrier ({\em transmitted} electrons), 
contribute to the current and noise.

It is advantageous to use as a variable in the equations, 
instead of the kinetic energy, the total energy 
$\epsilon=\hbar^2 k_x^2/(2m)-\Phi(x)$, where
$\Phi(x)\equiv q\varphi(x)-q\varphi_m$ is the mean potential
referenced to the minimum. 
By such a definition, $\Phi(x)>0$ in all the region (which is convenient 
for further consideration), whereas the potential energy $-\Phi(x)$ 
is negative.
Equation (\ref{vlasns}), for the stationary case ($\partial/\partial t$=0), 
in terms of these variables may be written as 
$(\partial/\partial x) F(x,\epsilon)$=0.
Its solution, being invariant on $x$, is expressed simply through 
the distribution function at the injecting contact $F_c$

\begin{equation} \label{fst}
F(\epsilon) = F_c(\epsilon+\Phi_c)\,\theta(\epsilon),
\end{equation}
where $\Phi_c\equiv\Phi(0)$ is the potential at the contact, 
and the Heaviside step function $\theta(\epsilon)$ establishes 
the lower bound for the transmitted electrons. 
The fluctuation $\delta F$ is found from linearization of Eq.\ (\ref{vlasns})
around the mean values. Equivalently, one may just perturb the steady-state
solution (\ref{fst}) as a compound function, and get

\begin{eqnarray} \label{df1}
\delta F(\epsilon) &=& \delta F_c(\epsilon+\Phi_c)\,\theta(\epsilon)
+ \frac{\partial F_c(\epsilon+\Phi_c)}{\partial\epsilon}
(\delta\epsilon+\delta\Phi_c) \, \theta(\epsilon) \nonumber\\
&&+ F_c(\epsilon+\Phi_c)\,\frac{\partial \theta(\epsilon)}{\partial \epsilon}\,
\delta\epsilon.
\end{eqnarray}
Taking into account that the perturbation of the total energy $\epsilon$
is related to the perturbation of the potential by 
$\delta\epsilon$=$-\delta\Phi_x$ and using the property
$\partial\theta(\epsilon)/\partial\epsilon$=$\delta(\epsilon)$,
one finally obtains \cite{remark3}

\begin{eqnarray} \label{df}
\delta F(\epsilon) &=& \delta F_c(\epsilon+\Phi_c)\,\theta(\epsilon)
- F_c(\Phi_c)\,\delta\Phi_x\,\delta(\epsilon)
\nonumber\\&&
+ \frac{\partial F_c(\epsilon+\Phi_c)}{\partial\epsilon} \, 
(\delta\Phi_c-\delta\Phi_x)\, \theta(\epsilon).
\end{eqnarray}
The self-consistent potential fluctuations are defined as 
$\delta\Phi_x\equiv q\delta\varphi(x)-q\delta\varphi_m$,
$\delta\Phi_c\equiv \delta\Phi_0$. 
This means that $\delta\Phi_x$ is measured in a frame referenced to 
the {\em fluctuating} potential minimum ($\delta\Phi_{x_m}$=0).
It is clear, that in such a consideration the contact potential and 
its fluctuation are related to the potential barrier height according to
$\Phi_c=qU_m$, $\delta\Phi_c=q\delta U_m$.

Equations (\ref{fst}) and (\ref{df}) should now be substituted into the Poisson
equations for $\Phi(x)$ and $\delta\Phi_x$, correspondingly, to find 
the self-consistent potential profile and its fluctuation.

\subsection{Steady state}

First, we find the mean electron density as a function of the potential 
$\Phi$ by integrating $F$ over the momentum $k_x$ and changing the variable 
of integration 
$dk_x=(\sqrt{2m}/\hbar)(d\epsilon/2\sqrt{\epsilon+\Phi})$, we obtain

\begin{eqnarray} \label{den}
N(\Phi)= \int_0^{\infty} F_c(\epsilon+\Phi_c)
\frac{d\epsilon}{2\sqrt{\epsilon+\Phi}}.
\end{eqnarray}
Then, we solve the Poisson equation
$d^2\Phi/dx^2=(q^2/\kappa) N(\Phi)$, subject to the boundary conditions
at the minimum $\Phi(x_m)$=0, and at the receiving contact
$\Phi_l\equiv \Phi(l)=q(U+U_m)$.
First integration leads to the electric-field distribution

\begin{eqnarray} \label{e}
E(\Phi) = - \frac{1}{q}\frac{d\Phi}{dx} =
- \sqrt{\frac{2q}{\kappa}}\sqrt{h(\Phi)},
\end{eqnarray}
where

\begin{eqnarray} \label{h}
&&h(\Phi)=  \int_0^{\Phi} N(\tilde{\Phi}) d\tilde{\Phi}
=  \int_0^{\infty} F_c(\epsilon+\Phi_c)
(\sqrt{\epsilon+\Phi}-\sqrt{\epsilon}) d\epsilon \nonumber\\
&&= {\cal F}_0 \sqrt{\Phi} - {\cal F}_1
+ \frac{{\cal F}_2}{2\sqrt{\Phi}} + O\left(\frac{1}{\Phi^{3/2}}\right),
\quad \Phi\to\infty, \\
&&{\cal F}_j(\Phi_c) =
\int_0^{\infty} F_c(\epsilon+\Phi_c) \epsilon^{j/2} d\epsilon,
\quad j=0,1,2,\dots. \label{fk}
\end{eqnarray}
The similar expansion for the electron density is given by
\begin{equation} \label{ninf}
N(\Phi) = \frac{d}{d\Phi} h(\Phi) =
\frac{{\cal F}_0}{2\sqrt{\Phi}} - \frac{{\cal F}_2}{4\Phi^{3/2}}
+ O\left(\frac{1}{\Phi^{5/2}}\right).
\end{equation}
Integration of Eq.\ (\ref{e}) with the expansion (\ref{h}) yields 
at $x_m\ll x \lesssim l$

\begin{equation} \label{etlch}
\Phi^{3/2}
\left[ 1 + \frac{3{\cal F}_1}{{\cal F}_0}\frac{1}{\sqrt{\Phi}}\right]
\approx \frac{9}{8} \frac{q^2 {\cal F}_0}{\kappa}  \, (x - x_m)^2.
\end{equation}
This equation, taken at $x$=$l$, may then be used to find 
the mean current 

\begin{eqnarray} \label{I}
&&I = \frac{qA}{\sqrt{2m}} \int_0^{\infty} F_c(\epsilon+\Phi_c) 
d\epsilon = \frac{qA}{\sqrt{2m}} {\cal F}_0 \nonumber\\ 
&\approx& \frac{4}{9}\kappa A \sqrt{\frac{2q}{m}}
\frac{(U+U_m)^{3/2}}{(l-x_m)^2}
\left[ 1 + \frac{3{\cal F}_1}{{\cal F}_0}\frac{1}{\sqrt{q(U+U_m)}} \right],
\end{eqnarray}
where $A$ is the cross-sectional area.
Here, the leading factor $\sim U^{3/2}$ 
(if one neglects $x_m$, $U_m$ with respect to $l$, $U$, respectively) 
is the Child current, which corresponds to what would be expected if
all the electrons are injected with zero initial velocity.
It is independent of the injection, but it is a function
of the applied bias $U$, the length $l$, and the parameters of the material
(the dielectric permittivity $\kappa$, the effective mass $m$).
The next-order term $\sim U$ contains information about the injection 
distribution function and gives the correction due to the spread of 
electron momenta at the minimum, since 
${\cal F}_1/{\cal F}_0=(\hbar/\sqrt{2m})\langle k_x^2 \rangle
/\langle k_x \rangle$, 
where we denote the average values at the minimum by angular brackets.
For the case of the Maxwellian injection, 
$F_c(\epsilon)\propto\exp(-\epsilon/k_BT)$, this ratio becomes 
${\cal F}_1/{\cal F}_0=\sqrt{\pi k_BT}/2$, and formula (\ref{I}) 
leads to the Langmuir formula for a vacuum diode. 
\cite{prb00,ziel54}

{}From Eq.\ (\ref{etlch}), one can get the asymptotic formula 
for the potential profile 
$\varphi^{3/2}(x)=\frac{9}{4}\sqrt{m/2q}(I/\kappa A)\,x^2$.
Substituting the Child current, one obtains the universal behavior
$\varphi(x)=U\,(x/l)^{4/3}$, at $x_m\ll x \leq l$, 
independently of the injection.
The other quantities of interest tend to the following distributions:
$E(x)=-\frac{4}{3}(U/l)(x/l)^{1/3}$,
$N(x)=\frac{4}{9}(\kappa U/q l^2)(x/l)^{-2/3}$.
It is seen, that the time-averaged quantities, such as the mean current and
the spatial profiles, asymptotically at high biases are nonsensitive 
to the injection distribution function. 
(Electrons coming to the receiving contact with the energies much higher
than their injecting energies forget about their initial spreading.)
The injection distribution gives just a small correction to 
the lower-order terms, which however may be essential at intermediate biases.
In contrast, the current noise is sensitive to the injection
distribution in the {\em leading-order terms}, which decrease with bias, 
as will be demonstrated below.

\subsection{Current fluctuations}

The current fluctuation is obtained by integrating over the energy 
the fluctuation of the distribution function (\ref{df})
\begin{eqnarray} \label{dI}
\delta I &=& \frac{qA}{\sqrt{2m}}
\int_0^{\infty} \delta F_c(\epsilon+\Phi_c) d \epsilon
- \frac{qA}{\sqrt{2m}} F_c(\Phi_c) \delta\Phi_c \nonumber\\
&\equiv& \int_0^{\infty} \delta I_c(\epsilon+\Phi_c) d \epsilon
+ \delta I_{\rm Coul}.
\end{eqnarray}
Here, $\delta I_c(\epsilon)$ is the partial injected current fluctuation 
in a unit of energy. The last term $\delta I_{\rm Coul}$, which is the current 
fluctuation caused by the long-range Coulomb interactions, may also be 
expressed more generally as $\delta I_{\rm Coul} = (\partial I/\partial 
\Phi_c) \delta \Phi_c = (\partial I/\partial U_m) \delta U_m$,
reflecting the modulation effect of the potential barrier fluctuations.
To find that term, we need to obtain $\delta\Phi_c$ as a function
of the injected fluctuations $\delta F_c$, by solving the Poisson equation.

Integrating Eq.\ (\ref{df}) over the momentum $k_x$, one gets
the electron-density fluctuation as a sum of two contributions,
$\delta N=\delta N^{inj}+\delta N^{ind}$,
where the injected part

\begin{eqnarray} \label{dninj}
\delta N^{inj}(\Phi) = \int_0^{\infty} \delta F_c(\epsilon+\Phi_c)
\frac{d\epsilon}{2\sqrt{\epsilon+\Phi}},
\end{eqnarray}
and the induced part

\begin{eqnarray} \label{dnind}
&&\delta N^{ind}(\Phi) =
(\delta\Phi_c-\delta\Phi_x)
\int_0^{\infty}\frac{\partial F_c(\epsilon+\Phi_c)}{\partial\epsilon}
\frac{d\epsilon}{2\sqrt{\epsilon+\Phi}}
\nonumber\\&&
- \delta\Phi_x  \frac{F_c(\Phi_c)}{2\sqrt{\Phi}}
=\frac{dN}{d\Phi} \delta\Phi_x - \left(\frac{dN}{d\Phi}
+ \frac{F_c(\Phi_c)}{2\sqrt{\Phi}} \right)\delta\Phi_c.
\end{eqnarray}
Substitution to the Poisson equation yields

\begin{eqnarray} \label{nhom}
\hat{L}\delta\Phi_x&\equiv& 
\left[\frac{d^2}{dx^2}- \frac{q^2}{\kappa} \frac{dN}{d\Phi}\right]
\delta\Phi_x \nonumber\\ 
&=& -\frac{q^2}{\kappa} \left(\frac{dN}{d\Phi} + 
\frac{F_c(\Phi_c)}{2\sqrt{\Phi}} \right)
\delta\Phi_c + \frac{q^2}{\kappa} \delta N^{inj}(\Phi).
\end{eqnarray}
By solving this equation with the boundary conditions $\delta\Phi_{x_m}$=0,
$\delta\Phi_l$=$\delta\Phi_c$ (see the appendix)
we find the Coulomb correlation term in the form

\begin{eqnarray} \label{dIcc}
\delta &&I_{\rm Coul} = \nonumber\\ &&
-\int_0^{\infty} \delta I_c(\epsilon+\Phi_c) 
\left[ 1-\frac{3}{\sqrt{qU}} 
\left(\sqrt{\epsilon} - \frac{N_m}{F_c(\Phi_c)} \right)\right] d\epsilon,
\end{eqnarray}
where $N_m$ is the electron density at the potential minimum.
Substitution of the found expression for $\delta I_{\rm Coul}$ into 
Eq.\ (\ref{dI}) for the total current fluctuation shows,
that the leading-order terms, which do not depend explicitly
on bias, are canceled, i.e., the injected current fluctuation is suppressed. 
The remaining contribution 

\begin{eqnarray} \label{dItot1}
\delta I = \frac{3}{\sqrt{qU}}
\int_0^{\infty} 
\left[\sqrt{\epsilon} - \frac{N_m}{F_c(\Phi_c)} \right] 
\delta I_c(\epsilon+\Phi_c) d\epsilon 
\end{eqnarray}
is $\propto U^{-1/2}$. 
We rewrite this expression in the form

\begin{eqnarray} \label{dItot}
\delta I = \int_{\Phi_c}^{\infty} \gamma(\varepsilon)
\delta I_c(\varepsilon) d\varepsilon,
\end{eqnarray}
in which the effect of the interactions is summarized by the quantity 
$\gamma(\varepsilon)$ determined by 

\begin{eqnarray} \label{gamma}
\gamma(\varepsilon) &=& \frac{3}{\sqrt{qU}} 
\left[ \sqrt{\varepsilon-\Phi_c} - \upsilon(\Phi_c) \right],
\end{eqnarray}
and the introduced energy $\varepsilon=\epsilon+\Phi_c$ corresponds to 
the (longitudinal) kinetic energy of electrons at the injecting contact. 
The constant $\upsilon$ in Eq.\ (\ref{gamma}) is the characteristic 
velocity given by 

\begin{eqnarray} \label{upsilon}
\upsilon(\Phi_c) &=& \frac{N_m}{F_c(\Phi_c)} \nonumber\\
&=& \frac{1}{F_c(\Phi_c)} \int_{\Phi_c}^{\infty} 
\left[- \frac{\partial F_c}{\partial \varepsilon}
\right] \sqrt{\varepsilon-\Phi_c} d\varepsilon. \label{vv}
\end{eqnarray}
The main result, which follows from the derived expression 
(\ref{gamma}), is that $\gamma(\varepsilon)$ is a {\em decreasing function 
of the applied bias $U$}.
With higher bias, a larger suppression of the current fluctuations is
expected.
Another important conclusion is that the suppression effect is different for 
different injection shapes $F_c(\varepsilon)$. 
The dependence on $F_c(\varepsilon)$ is summarized by the characteristic 
velocity $\upsilon$ determined by Eq.\ (\ref{upsilon}).
Note, that the function $\gamma(\varepsilon)$ has a meaning of the current 
fluctuation transfer function, \cite{prb00} and in general may be as positive, 
as negative depending on the particular energy $\varepsilon$.
In the absence of correlations,
$\gamma^{uncor}(\varepsilon)=\theta(\varepsilon-qU_m)$,
that means the fluctuations of all energies above the barrier height $qU_m$
are equally transmitted. 

Having found the current fluctuation $\delta I$ expressed through
the injected current fluctuations $\delta I_c(\epsilon)$,
the current-noise spectral density may then be obtained
from Eq.\ (\ref{dItot}) as

\begin{equation} \label{si}
S_I \Delta f = 
\int_{\Phi_c}^{\infty} \int_{\Phi_c}^{\infty}  
\gamma(\varepsilon)\gamma(\varepsilon')
\langle\delta I_c(\varepsilon)\delta I_c(\varepsilon')\rangle 
d\varepsilon d\varepsilon'.
\end{equation}
Here, $\Delta f$ is the frequency bandwidth 
(we assume the low-frequency limit), and in such a presentation the function 
$\gamma(\varepsilon)$ plays the role 
of the {\em energy resolved shot-noise-suppression factor}. 

The incoming electrons may be correlated in energy {\em a priori} due to 
the properties of an emitter. 
In general case of non-Poissonian injection, one can define 
the shot-noise-suppression factor due to a pure Coulomb suppression by

\begin{equation} \label{sup}
\Gamma_C =
\frac{\int_{\Phi_c}^{\infty} \int_{\Phi_c}^{\infty} 
\gamma(\varepsilon)\gamma(\varepsilon')
\langle\delta I_c(\varepsilon)\delta I_c(\varepsilon')\rangle 
d\varepsilon d\varepsilon'}
{\int_{\Phi_c}^{\infty}\int_{\Phi_c}^{\infty}
 \langle\delta I_c(\varepsilon)\delta I_c(\varepsilon')\rangle 
d\varepsilon d\varepsilon'}, 
\end{equation}
which can be easily found when the properties of injected carriers are given.

For the particular case when the injected carriers of different energies are 
uncorrelated,

\begin{equation} \label{kuncor}
\langle\delta I_c(\varepsilon)\delta I_c(\varepsilon')\rangle = 
K(\varepsilon)(\Delta f)\delta(\varepsilon-\varepsilon'),
\end{equation}
the shot-noise suppression factor (\ref{sup}) is simplified to

\begin{equation} \label{supuncor}
\Gamma_C =
\frac{\int_{\Phi_c}^{\infty} 
\gamma^2(\varepsilon) K(\varepsilon) d\varepsilon}
{\int_{\Phi_c}^{\infty} K(\varepsilon) d\varepsilon}.
\end{equation}

Furthermore, for the Poissonian injection, the property of the kernel $K$ is
such that
$K(\varepsilon) \propto I_c(\varepsilon) \propto F_c(\varepsilon)$.
Hence, one can find

\begin{equation} \label{suppois}
\Gamma_{\rm Poisson}=
\frac{\int_{\Phi_c}^{\infty} 
\gamma^2(\varepsilon) F_c(\varepsilon) d\varepsilon}
{\int_{\Phi_c}^{\infty} F_c(\varepsilon) d\varepsilon}
\to \frac{S_I}{2qI}.
\end{equation}

Depending on the injection, one of the expressions (\ref{sup}), 
(\ref{supuncor}), (\ref{suppois}) can be used together with the function 
$\gamma(\varepsilon)$ given by Eq.\ (\ref{gamma}) to evaluate the shot-noise 
suppression-level in ballistic space-charge-limited conductors under 
the action of Coulomb interactions.
Note that the formula for $\gamma(\varepsilon)$ is valid for any given 
energy distribution and statistical properties of the injected carriers
under the condition of a high bias, that is $U \gg U_m$ and $U$ much larger 
than the characteristic energy spreading of injected electrons.
The upper bound for the bias is however restricted by the condition 
of the existence of the potential barrier $U < U_{cr}$ 
(space-charge-limited transport).
Both conditions may be fulfilled simultaneously under a sufficiently strong 
screening, i.e., the length of the conductor should be much larger than 
the characteristic screening length. \cite{prb00}

\section{Examples}

To illustrate the implementation of the results, we consider some examples.
For the Maxwell-Boltzmann (MB) injection distribution
(nondegenerate equilibrium electron gas is injected)
we obtain $\upsilon=\sqrt{\pi k_BT}/2$, i.e., it only depends
on the temperature of the injected electrons, but otherwise is independent
of the material parameters, since its dependence on the barrier height 
is canceled out.
For this case Eq.\ (\ref{gamma}) gives

\begin{equation} \label{gammaxw}
\gamma_{MB}(\varepsilon) = 3\sqrt{\frac{k_B T}{qU}}
\left( \sqrt{\frac{\varepsilon-\Phi_c}{k_B T}} - \frac{\sqrt{\pi}}{2}\right), 
\end{equation}
which coincides with the formula derived by North. \cite{prb00,north40} 
The corresponding shot-noise-suppression factor follows from 
Eq.\ (\ref{suppois})

\begin{equation} \label{Gammaxw}
\Gamma_{MB} = 9 \left( 1 - \frac{\pi}{4} \right) \frac{k_BT}{qU}.
\end{equation}
For a quantitative estimation consider the heterodiode with 
GaAs contacts and an Al$_{0.05}$Ga$_{0.95}$As ballistic base. \cite{apl99}
For the contact doping $4\times 10^{16} {\rm cm}^{-3}$ at $T$=50 K, 
we obtain the injected electron density about 
$7.25\times 10^{14} {\rm cm}^{-3}$ which corresponds to the 
Debye screening length $L_D$=46 nm. 
Then for the $1.5 \mu$-length diode and $U\approx 45 k_BT/q$,
the noise-suppression level estimated from the exact solutions \cite{apl99}
gives $\Gamma_{MB}\approx 0.04$, which is close to 
the value calculated from the asymptotic formula (\ref{Gammaxw}).

Now we shall demonstrate, that the shot-noise suppression level may be 
achieved even deeper than that given by Eq.\ (\ref{Gammaxw}) for 
the MB case, without involving any other correlations (like the Pauli 
exclusion principle) in addition to the Coulomb correlations. 
The higher suppression may be achieved by modifying the energy profile
for the injected carriers. 
Consider the heterodiode under the same set of parameters considered above, 
in which, in addition to the Maxwell-Boltzmann injection,
nonequilibrium carriers are injected from a specially designed emitter, 
so that the injected distribution function has an additional 
peak at the energy $\varepsilon_0$ [see inset of Fig.\ 1(b)].
According to our theory, these additional electrons do not change
the current-voltage characteristics much. 
Its asymptotic behavior is again the Child law. 
However, the noise properties change significantly depending on 
the parameters of the electron-energy peak, its magnitude, position, etc.
In particular, the noise-suppression level may be obtained lower or higher
than the MB shot noise by simply shifting the position of the peak
(of the emitter) in respect to the potential barrier.

Let us assume that the width of the peak is narrow on the scale of 
the temperature $T$. For simplicity, we model it first
by a $\delta$ function (monoenergetic electrons)

\begin{equation} \label{distdelt}
F_c(\varepsilon) \propto e^{-\varepsilon/k_B T} 
+ \tilde{\alpha}\, k_B T\, \delta(\varepsilon-\varepsilon_0)
\end{equation}
The injected carriers, from both the Maxwellian tail and the peak, 
are assumed to be uncorrelated, so that Eq.\ (\ref{suppois})
can be applied.
Thus, for the distribution (\ref{distdelt}) one gets the (normalized) 
characteristic velocity (\ref{vv}) as

\begin{eqnarray} \label{upsdelt}
w \equiv \frac{\upsilon}{\sqrt{k_BT}} = \frac{\sqrt{\pi}}{2} \left(
1 + \frac{\alpha}{\sqrt{\pi\xi}} \right),
\end{eqnarray}
where $\alpha=\tilde{\alpha}e^{\Phi_c}$ is the ratio between the two currents:
from the $\delta$ peak and from the MB exponential tail, and 
$\xi=(\varepsilon_0-\Phi_c)/(k_BT)$ is the dimensionless position of the peak.
The shot-noise-suppression factor is then obtained as

\begin{equation} \label{Gamdelt}
\Gamma = 9\, \frac{k_BT}{qU}\,
\frac{1-w\sqrt{\pi}+w^2 + \alpha (\sqrt{\xi}-w)^2 }{1+\alpha}.
\end{equation}
\begin{figure}
\narrowtext
\epsfxsize=8.0cm
\epsfbox{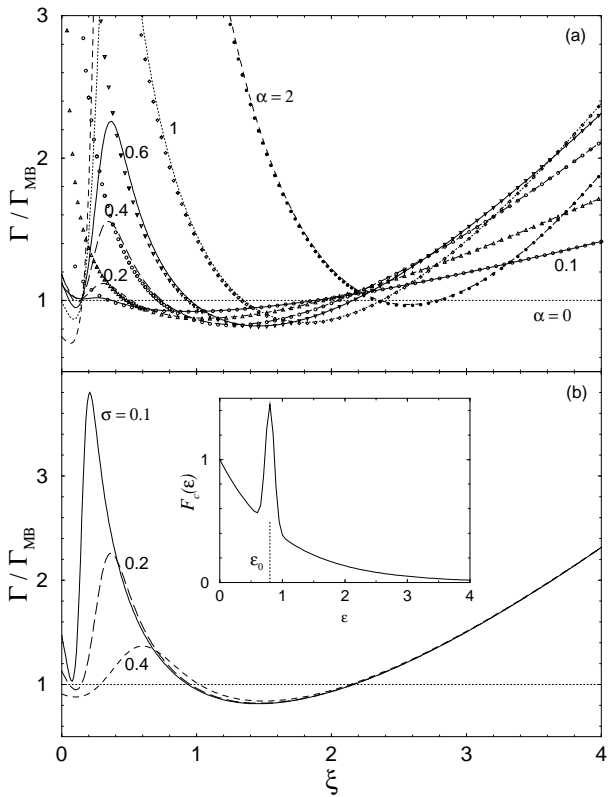}
\protect\vspace{0.5cm}
\caption{
The shot-noise-suppression level $\Gamma$ (caused by Coulomb interactions) 
for the Maxwell-Boltzmann injection with an additional peak at 
$\varepsilon$=$\varepsilon_0$ (shown in the inset) 
with respect to the case when no peak is present. 
The ratio $\Gamma/\Gamma_{MB}$ is shown as a function of the peak position 
$\xi=(\varepsilon_0-\Phi_c)/(k_BT)$.
(a) the peak parameter $\alpha$ is varied. 
The results are compared for two different shapes of the peak:
$\delta$ function given by Eq.\ (\ref{distdelt}) (symbols); 
Gaussian function given by Eq.\ (\ref{dist}) for $\sigma$=0.2 (lines).
(b) the Gaussian peak case: the width of the peak $\sigma$ is varied, 
while $\alpha$=0.6 is fixed.}
\label{f1}
\end{figure}

\noindent
In the limit when the electron-energy peak vanishes, $\alpha\to 0$,
formula (\ref{Gamdelt}) is reduced to the suppression factor (\ref{Gammaxw})
for the MB injection.
We have studied how $\Gamma$ deviates from $\Gamma_{MB}$ when the peak current
$\alpha$ and the peak position $\xi$ are varied. 
The results are illustrated in Fig.\ 1(a).
The dependence of $\Gamma/\Gamma_{MB}$ on $\xi$ 
was found to be nonmonotonic displaying a minimum.
In some range of $\xi$ around the minimum,
$\Gamma/\Gamma_{MB} < 1$,
that means the additional electron-energy peak at the injecting contact
results in a less noisy transmission, than in the case of its absence.
The minimal noise is observed for $\alpha\approx 0.6$, for which we find
$\Gamma/\Gamma_{MB}\approx 0.814$ 
at $\xi_{min} \approx 1.45$.
As follows from Fig.\ 1(a),
the most effective noise suppression occurs when the peak is about 1--2 $k_BT$
above the barrier. When it is higher in energy, or too close to the 
barrier position ($\varepsilon_0-\Phi_c \lesssim k_BT$), 
the noise is enhanced in respect to the MB case 
(although it may still be below the Poissonian value).

The analysis for other shapes of the peak shows that the results
are similar to those for the $\delta$ peak. 
As an example, we present here the results for the case when the peak 
is modeled by the Gaussian distribution function

\begin{equation} \label{dist}
F_c(\varepsilon) \propto e^{-\varepsilon/k_B T} 
+ \varrho\, e^{-(\varepsilon-\varepsilon_0)^2/(\sigma k_B T)^2},
\end{equation}
where the factor $\varrho$ is defined by 
$\varrho=2\tilde{\alpha}/
\{ \sigma\sqrt{\pi}[1+{\rm erf}(\varepsilon_0/\sigma k_B T)] \}$.
By such a definition, the parameter
$\alpha=\tilde{\alpha}e^{\Phi_c}$ gives again, as in the previous case, 
the ratio between the current originated from the Gaussian peak and that 
from the MB tail.
A comparison between the two cases is presented in Fig.\ 1(a).
It is seen, that at high values of $\xi$ the results for the noise 
suppression for both cases of the Gaussian and $\delta$ peak coincide.
It can be shown, that this occurs at $\xi\gtrsim 5\sigma$.
Hence, when the peak width $\sigma < (\xi_{min}/5)\approx 0.3$,
the minimal noise occurs at the same peak position $\xi_{min}$, 
independently of the value of $\sigma$.
The information on the peak width is presented, however, 
in the noise-suppression curves at low values of $\xi$.
While $\Gamma$ for the $\delta$ peak case diverges at $\xi\to 0$,
due to a singularity of the $\delta$ function, the noise suppression factor
for the Gaussian-peak case exhibits a local maximum 
[see Figs.\ 1(a) and 1(b)].
The magnitude of this noise enhancement (in respect to the MB case) depends 
on $\sigma$: the narrower the peak, the larger the noise enhancement and 
the closer is the location of the maximum to the potential barrier energy
[see Fig.\ 1(b)].

Summarizing this example, to observe the lower noise level for nondegenerate 
ballistic electrons, the additional (to the MB tail) electrons should be 
injected with the energy about 1.45 $k_BT$ above the potential barrier. 
This value is independent of the energy spreading of the ``peak'' electrons
once the latter is less than 0.3 $k_BT$. 
The optimal ratio between the current from the ``peak'' electrons and 
the MB electrons is about 0.6.

It is seen, that the shot noise contains important information on: 
(i) the injection energy profile, and (ii) the parameters of the injected 
space charge, such as the potential barrier height 
and the electron density $N_m$ at the barrier position. 
Therefore, noise measurements may be used as a tool to study those
characteristics. 

\section{Summary}

In conclusion, we have presented a self-consistent theory of transport and 
current noise in two-terminal ballistic space-charge-limited conductors 
under the action of Coulomb interactions.
We have derived the analytical formulas that account for the non-Poissonian 
injection with arbitrary distribution function and correlation properties 
of injected electrons, and these may be used to estimate: 
(i) the mean current beyond the Child approximation with a next-order term
specific of the injection distribution function;
(ii) the current-noise spectral density under the action of Coulomb 
interactions, which depends in the leading-order terms on the injection
distribution function and decreases with bias;
(iii) the noise-suppression factor in respect to the injected
non-Poissonian electron flow. \cite{remark1}

The obtained analytical formula for the energy-resolved shot-noise 
suppression may suggest a new field of investigation on the optimization of
the injection energy profile to achieve the desired noise suppression level. 
The presented examples clearly show, that the noise-suppression level may be 
controlled by monitoring the injection energy profile.

The sensitivity of the noise-suppression level to the injection parameters 
opens up new perspectives in shot-noise measurements as a tool
not only to identify the ballistic transport in mesoscopic conductors,
but also to reveal an important information on the injection energy profile
and the level of Coulomb interactions in the structure.
Experiments have succeeded recently in observing shot noise in ballistic
quantum point contacts \cite{reznikov95,kumar96} and some other mesoscopic 
systems (see, e.g., Refs.\ \onlinecite{steinbach96,schoelkopf97,jehl99}).
We believe, that it would be similarly possible to measure the shot noise
in space-charge limited ballistic conductors.

Additionally, it is important to emphasize the difference between
the asymptotic behavior of the shot noise in diffusive and ballistic systems 
under the presence of a space charge.
In the former case the noise-suppression level is limited by the constant,
specific of the dominating scattering mechanism,
\cite{gonzalez98a,gonzalez99,beenakker99,schomerus99}
while in the latter the suppression may be arbitrarily strong,
which may be important from the point of view of possible applications.

\acknowledgements

This work was partially supported by the Generalitat de Catalunya,
Spain, and the NATO linkage grant HTECH.LG 974610.

\appendix
\section{Derivation of the self-consistent potential fluctuations}

The second-order differential equation (\ref{nhom}) with spatially dependent 
coefficients can be solved explicitly for $\delta\Phi_x$.
\cite{prb00,apl97}
Here, we need just the value of $\delta\Phi_c$, which has entered explicitly 
into the nonhomogeneous part and can be obtained 
by applying the Green's theorem for the self-adjoint operator $\hat{L}$

\begin{eqnarray} \label{green}
\int_{x_m}^l [u(x) \hat{L}\delta\Phi_x &-& \delta\Phi_x\hat{L}u(x)] dx
\nonumber\\
&=& \left. \left( u(x)\frac{d\delta\Phi}{dx} - \delta\Phi_x \frac{du}{dx}
\right)\right|_{x_m}^l.
\end{eqnarray}
It is convenient to chose the function $u(x)$ as a solution of the 
homogeneous equation $\hat{L}u(x)$=0 satisfying the boundary condition 
$u(l)$=0. This gives

\begin{eqnarray} \label{green2}
-\delta\Phi_c \frac{q^2}{\kappa} \int_{x_m}^l u &&
\left(\frac{dN}{d\Phi} + \frac{F_c(\Phi_c)}{2\sqrt{\Phi}} \right) dx
+ \frac{q^2}{\kappa} \int_{x_m}^l u \, \delta N^{inj} dx
\nonumber\\ &&
= -u'(l)\delta\Phi_c - u(x_m) \delta\Phi'_{x_m},
\end{eqnarray}
where prime stands for the derivative on $x$.
It can be shown, that at large $U$, both terms in the right-hand side 
of Eq.\ (\ref{green2}) may be neglected.
Indeed, \cite{remark4}
$u'(l)$=1/$E(l)$=$O(\Phi_l^{-1/4})\to 0$, at $\Phi_l\to\infty$.
The term $u(x_m) \delta\Phi'_{x_m}$ may be evaluated from the matching
with the expression similar to Eq.\ (\ref{green2}) for the adjacent region 
$0<x<x_m$. It occurs to be $O(1)$ at $\Phi_l\to\infty$, and hence
gives negligible contribution in respect to the leading terms 
$O(\Phi_l^{3/4})$ (see below).
Changing the variable of integration $dx$=$-d\Phi/(qE)$ , one gets

\begin{eqnarray} \label{green3}
\delta\Phi_c
\int_0^{\Phi_l} \frac{u}{E}
\left(\frac{dN}{d\Phi} + \frac{F_c(\Phi_c)}{2\sqrt{\Phi}} \right) d\Phi
= \int_0^{\Phi_l} \frac{u}{E} \, \delta N^{inj} \, d\Phi.
\end{eqnarray}
In this equation the integrals may be integrated by parts in a similar way

\begin{eqnarray} \label{parts}
\int_0^{\Phi_l} \frac{u}{E} \frac{d G_i}{d\Phi} \, d\Phi
&=& \left.\left( \frac{u}{E} G_i \right) \right|_0^{\Phi_l}
- \int_0^{\Phi_l} G_i \,\frac{d}{d\Phi} \left(\frac{u}{E}\right) d\Phi \\
&=& \frac{1}{q} \int_0^{\Phi_l} \frac{G_i}{E^3} \, d\Phi, 
\qquad i=1,2, \nonumber
\end{eqnarray}
with 

\begin{eqnarray} \label{g12}
G_1(\Phi) &=& N(\Phi) - N(0) + F_c(\Phi_c)\sqrt{\Phi}, \\
G_2(\Phi) &=& \int_0^{\infty} \delta F_c(\epsilon+\Phi_c)
(\sqrt{\epsilon+\Phi}-\sqrt{\epsilon}) \,d\epsilon.
\end{eqnarray}
Notice, that the first term in right-hand side of Eq.\ (\ref{parts}) is zero,
since at the upper limit $u(\Phi_l)$=0, and at the lower limit we have
$G_i(\Phi)\sim \Phi$, $E(\Phi)\sim \sqrt{\Phi}$ at $\Phi\to 0$, and
$u(0)=\kappa/[q N(0)]$ is finite.
In the second integral of Eq.\ (\ref{parts}) we have used \cite{remark4}

\begin{equation}
\frac{d}{d\Phi} \left(\frac{u}{E}\right)
= - \frac{1}{qE} \frac{d}{dx} \left(\frac{u}{E}\right)
= - \frac{1}{qE^3}.
\end{equation}
Thus, Eq.\ (\ref{green3}) becomes

\begin{eqnarray} \label{green4}
\delta\Phi_c&&
\int_0^{\Phi_l}
\frac{N_m - N(\Phi) - F_c(\Phi_c)\sqrt{\Phi}}{E^3(\Phi)} d\Phi
\nonumber\\ && = 
\int_0^{\infty} d\epsilon \delta F_c(\epsilon+\Phi_c)
\int_0^{\Phi_l} d\Phi 
\frac{\sqrt{\epsilon+\Phi}-\sqrt{\epsilon}}{E^3(\Phi)},
\end{eqnarray}
where $N_m\equiv N(\Phi$=0) is the electron density at the potential minimum.
At the high-bias limit $\Phi_l\to\infty$, by using 
Eqs.\ (\ref{e})--(\ref{ninf}), one obtains

\begin{eqnarray} \label{ex1}
\int_0^{\Phi_l} \frac{N_m - N(\Phi) - F_c(\Phi_c)\sqrt{\Phi}}{E^3(\Phi)} 
d\Phi = \frac{4}{3} \Phi_l^{3/4} F_c(\Phi_c) \nonumber\\  \times
\left[ 
1+3 \left(\frac{3{\cal F}_1}{2{\cal F}_0}-\frac{N_m}{F_c(\Phi_c)}\right) 
\Phi_l^{-1/2} + O(\Phi_l^{-1}) \right],
\end{eqnarray}
\begin{eqnarray} \label{ex2}
&&\int_0^{\Phi_l} \frac{\sqrt{\epsilon+\Phi}-\sqrt{\epsilon}}{E^3(\Phi)} 
d\Phi = \frac{4}{3} \Phi_l^{3/4} \nonumber\\ && \hspace*{1cm} \times 
\left[ 1+3 \left(\frac{3{\cal F}_1}{2{\cal F}_0}-\sqrt{\epsilon} \right) 
\Phi_l^{-1/2} + O(\Phi_l^{-1}) \right].
\end{eqnarray}
In the latter expansion it is assumed, that the range of valuable injection 
energies is much less than the applied bias, $\epsilon\ll\Phi_l$.
Substituting these expansions into Eq. (\ref{green4}), one obtains 

\begin{eqnarray} \label{dfcc}
&& F_c(\Phi_c)\delta\Phi_c = \nonumber\\ &&
\int_0^{\infty} \delta F_c(\epsilon+\Phi_c)
\left[ 1-\frac{3}{\sqrt{\Phi_l}}
\left(\sqrt{\epsilon} - \frac{N_m}{F_c(\Phi_c)} \right)\right] d\epsilon,
\end{eqnarray}
which is used to find the Coulomb correlation term $\delta I_{\rm Coul}$
in Eq.\ (\ref{dI}).

\end{multicols}

\begin{references}
\bibitem{remark2}
The case $U>U_{cr}$ is not interesting, since the transport is no longer 
space-charge limited, and the noise is no longer suppressed by Coulomb 
correlations.
\bibitem{remark3}
One should not confuse the notations $\delta\epsilon$ and 
$\delta(\epsilon)$. The former means a (small) variation in the total energy 
$\epsilon$, while the latter is the Dirac $\delta$ function on the argument
$\epsilon$.
\bibitem{remark4}
The function $u(x)$ can be found through the second independent solution 
$E(x)$ of the homogeneous equation $\hat{L}E(x)$=0,
with the boundary condition $E(x_m)$=0.
Here, however, we do not need an explicit formula for $u$, employing just
the relation $Eu'-E'u$=$W$=1, where $W$ is the Wronskian 
(Ref.\ \onlinecite{prb00}).
\end{references}
\end{document}